\documentclass[useAMS,usenatbib]{mn2e}
\usepackage{graphics}
 \usepackage{graphicx}
 \usepackage{float}
\usepackage{placeins}
\usepackage{color}

\title[Survival of impactor on highly porous targets]{Survival of the impactor during hypervelocity collisions II:\\ An analogue for high porosity targets.}
\author[C. Avdellidou et al.]{C. Avdellidou$^{1}$\thanks{E-mail:
ca332@star.kent.ac.uk}, M.C. Price$^{1}$, M. Delbo$^{2}$, and  M.J. Cole$^{1}$\\
$^{1}$Centre for Astrophysics and Planetary Science, School of Physical Sciences, University of Kent, Canterbury, CT2 7NH, UK\\
$^{2}$Laboratoire Lagrange, Universit\'e C\^ote d'Azur, Observatoire de la C\^ote d'Azur, CNRS, Blvd de l'Observatoire, CS 34229,\\ 06304 Nice cedex 4, France\\}

\begin{document}

\date{Accepted . Received }

\pagerange{\pageref{firstpage}--\pageref{lastpage}} \pubyear{2002}

\maketitle

\label{firstpage}

\begin{abstract}
We investigated how a target's porosity affects the outcome of a collision, with respect to the impactor's fate. Laboratory impact experiments using peridot projectiles were performed at a speed range between 0.3 and 3.0 km/s, onto high porosity water-ice (40\%) and fine-grained calcium carbonate (70\%) targets.
We report that the amount of implanted material in the target body increases with increasing target's porosity, while the size frequency distribution of the projectile's ejecta fragments becomes steeper. A supplementary Raman study showed no sign of change of the Raman spectra of the recovered olivine projectile fragments indicate minimal physical change. 
\end{abstract}


\section{Introduction}
\label{intro}
The discovery of multi-lithology meteorites, such as the Almahata Sitta \citep[from the asteroid 2008 TC$_{3}$,][]{2009Natur.458..485J, 2010MAPS...45.1638B} and Benesov \citep{2014benesov}, and the identified exogenic material on asteroids, e.g. on Vesta \citep{2012Icar..221..544R,palomba2014}, Itokawa \citep{fujiwara2006,hirata2011} and Lutetia \citep{belskaya2010,barucci2012,schroder2015}, raises fundamental questions:
what is the possibility of forming these objects by collisions between bodies of different composition? Are asteroids with mixed mineralogies more abundant than it was previously thought? However, the formation mechanism(s) for these bodies remain a mystery \citep{2014horstmann}. If the formation mechanism via impacts of bodies with diverse compositions is effective, the discovery of impactor residues on a target could reveal details about the impact history of the body and/or the impactors populations. Can these bodies be formed in the current asteroid belt or they are formed 4.5 Ga when small bodies where more numerous and impacts more frequent? When considering the implantation of exogenic material on an asteroid, we currently assume only very low-speed collisions \citep{2012MNRAS.424..508G}. This is due to the preconception that, during hypervelocity impacts (a few km/s), the projectile is totally vaporised \citep[e.g.][]{ammannito2013}. However, recent work by \cite{me2016} shows that this is not necessarily the case. Additionally,
\cite{Dalynew,2016Daly} used aluminum and basalt projectiles  which were fired onto pumice and highly porous water-ice targets,  simulating the implantation of an impactor's material onto the vestan and Ceres' regolith. They found that material can be deposited via impacts, with the amount decreasing with impact angle, using impact speeds in a narrow regime between 4.4 and 4.9 km/s.
The main question that is addressed here is how much of the impactor's material is embedded on/into the target as a function of its porosity.

\section{Experiments}
\label{experiments}

We carried out low- and hyper-velocity impact experiments (0.30--3.0 km/s) similar to that of \cite{me2016} (hereafter Run\#1); with the main difference that we used targets with moderate to high porosity. Both experiments were performed by using the horizontal two-stage Light Gas Gun (LGG) of the University of Kent \citep{1999MeScT..10...41B} and two different setups were built to capture the projectile's ejecta for further examination (see Fig.~1). 

%We fired olivine projectiles in low- and hyper-velocity impacts (0.30--3.0 km/s) onto highly porous water-ice ($\sim$40\%) and high-porosity calcium carbonate (CaCO$_{3}$) powder ($\sim$70\%) targets.
Our aims were to: (a) study the fragmentation of the projectile, (b) derive its energy density at the catastrophic disruption threshold, Q*$_{\mathrm{im}}$, (c) measure the size frequency distributions (SFDs) of the projectiles' fragments in the ejecta, (d) estimate the implanted mass in the target, and (e) examine the physical state of the surviving projectile fragments.
%Expanding previous experiments \citep{me2016}, hereafter we refer as Run\#1, we used a combination of  materials, for both projectiles and targets, allowing the materials to be separated from each other, leaving the projectile's post-impact mass clear for further study. The aim is to: (a) study the fragmentation of the projectile, (b) derive its energy density at the catastrophic disruption threshold, Q*$_{\mathrm{im}}$, (c) measure the size frequency distributions (SFDs) of the projectiles' fragments in ejecta, (d) estimate the implanted mass in the target, and (e) examine the state of the surviving fragments.
%In this work we continue the investigation of the fate of the projectile during low- and hypervelocity impacts (0.30--3.0 km/s) by firing olivine onto highly porous water-ice ($\sim$40\%) and high-porosity calcium carbonate (CaCO$_{3}$) powder ($\sim$70\%) targets. A part of the experiments was analysed using the methods described by \cite{me2016}. 

\subsection{Materials and setup}
\label{expmethod}

We used olivine projectiles because, together with pyroxene, olivine is one of the most common minerals in the Solar System and is found in asteroids \citep{propertiesmeteorites,2002aste.conf..183G,2011Sci...333.1113N}, comets \cite[e.g. {\it Stardust Mission,}][]{2006Sci...314.1735Z} and planets. %In addition, there are strong indications that olivine has been identified to exo-planetary systems (e.g. $\beta$ Pictoris, \cite{2012Natur.490...74D}). 
The projectiles were 3 mm peridots, high purity Mg-rich olivine. All projectiles were examined by Raman spectroscopy at IR (784 nm) and Energy-dispersive X-ray spectroscopy (EDX), revealing homogeneity in the same projectile and identical composition between them, a very important aspect for the reproducibility of our experiments. 
We used two different types of target and ejecta collection setups:
(a) In Run\#2 water-ice targets with porosity 35-40\% were prepared by spraying high purity water into liquid nitrogen. The ice grains had a range of sizes from sub-mm to a few mm, comparable with the projectile's size. After each shot, in order to recover the projectile's fragments, the icy target and the ice from the ejecta collection setup were left to melt. 
(b) In Run\#3 CaCO$_{3}$ powder with 70\% porosity simulating a regolith-like surface. Using this target we followed a slightly different procedure, as in order to collect the projectile's fragments, we had to dissolve it in nitric acid to leave the olivine projectile fragments behind. 
\begin{figure}
\includegraphics[width=1.\columnwidth]{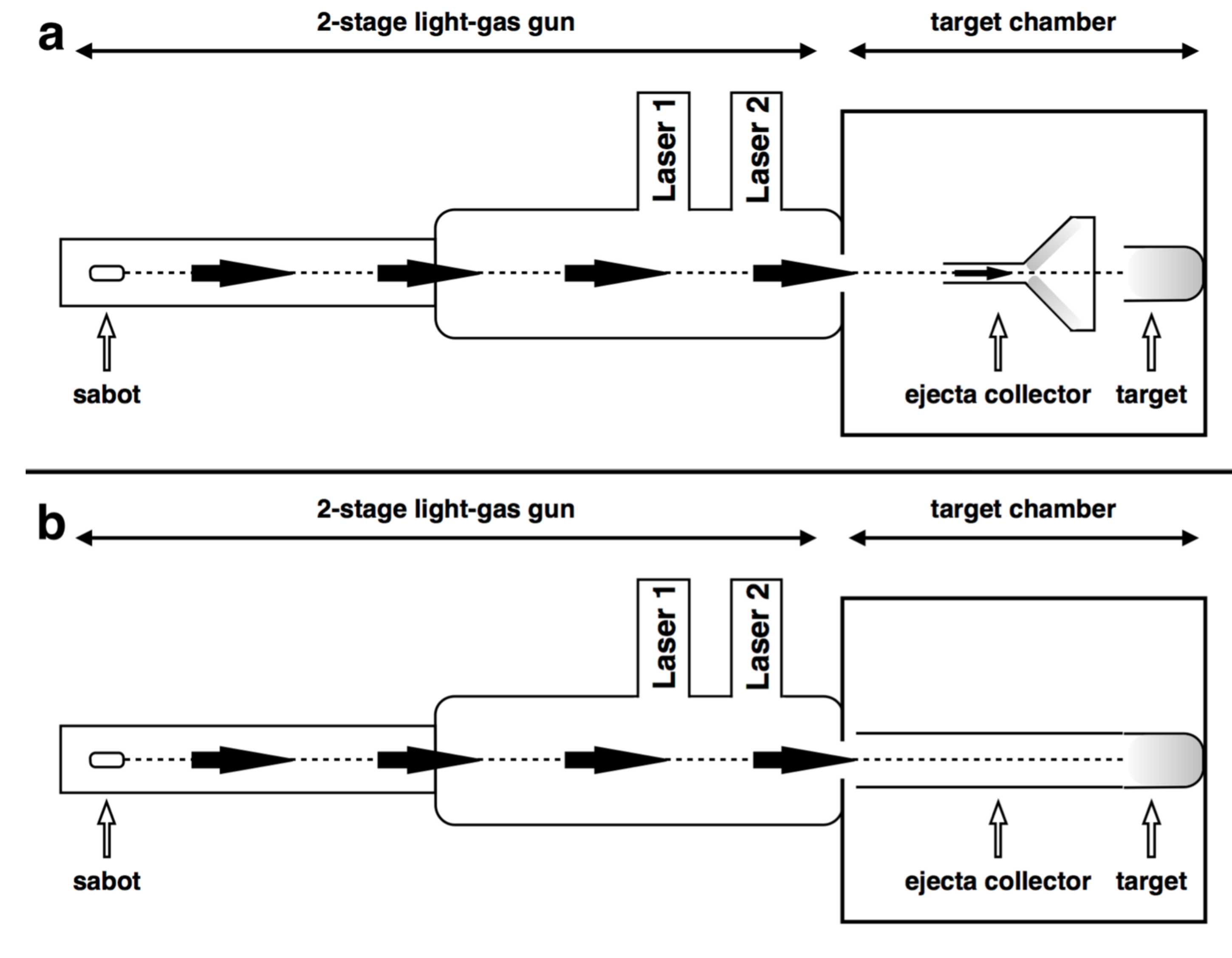}
\caption{The experimental setup, used for the icy (a) and regolith-like (b) targets, showing the projectile, which was placed in a sabot, inside the two-stage LGG, and the configuration of the target chamber. The projectile impacts the target at 0$^{\circ}$ with respect to its trajectory (dashed line). The ejecta collection funnel (a), the same used for Run\#1 with low porosity water-ice targets as described in \protect\cite{me2016}, was aligned with the flight path of the projectile and the centre of the target. It contained water-ice layers in order to collect the projectile's debris after the impact. For the regolith shots (b) a plastic tube was used to capture all the ejecta with no internal ice coating. In that way the loss of projectile fragments was minimised. We acknowledged that the ejecta collecting systems could possibly lead to a secondary fragmentation, but this is considered minimal due to the low ejection speeds that were observed using porous targets.}
\label{setup}
\end{figure}

The grainy material was held in horizontal place with no help of another layer (e.g. membrane), but only due to the compaction (Run\#3) and the air condensation due to low temperature ($-130${$^{\circ}$}C) that were kept in prior to each shot (Run\#2 and Run\#3).
Each target's temperature during impact was $-50${$^{\circ}$}C and the target chamber's pressure was set to 50 mbar for all experiments. 
From both experiments the water-ice melt or CaCO$_{3}$ solution in nitric acid were filtered through PTFE (polytetrafluorethylene) filters with 0.1 and 5 $\umu$m pore-size for the target and ejecta liquids respectively. As experiments by \cite{2008LPI....39.2045B} and \cite{Dalynew,2016Daly}, have shown that the largest portion of the impactors mass is kept on the target at 90$^{\circ}$ with respect to the target's surface, we carried out our experiments using the same configuration. 

\section{Results and Discussion}
\label{results}

\subsection{Impactor's fragmentation}
\label{state}

After collection of the impactor's fragments from the ejecta, their sizes were measured using the same technique as described in \cite{me2016}.
The olivine fragments were identified with Energy-dispersive Spectroscopy (EDX maps) as forsterite gives a strong signal in Mg.
The Mg maps were processed with SOURCE EXTRACTOR (SEXTRACTOR), an astronomical software which specialises in photometry and extraction of the light of irregular sources in dense fields \citep{1996A&AS..117..393B}.
 
\begin{figure}
\includegraphics[width=1.\columnwidth]{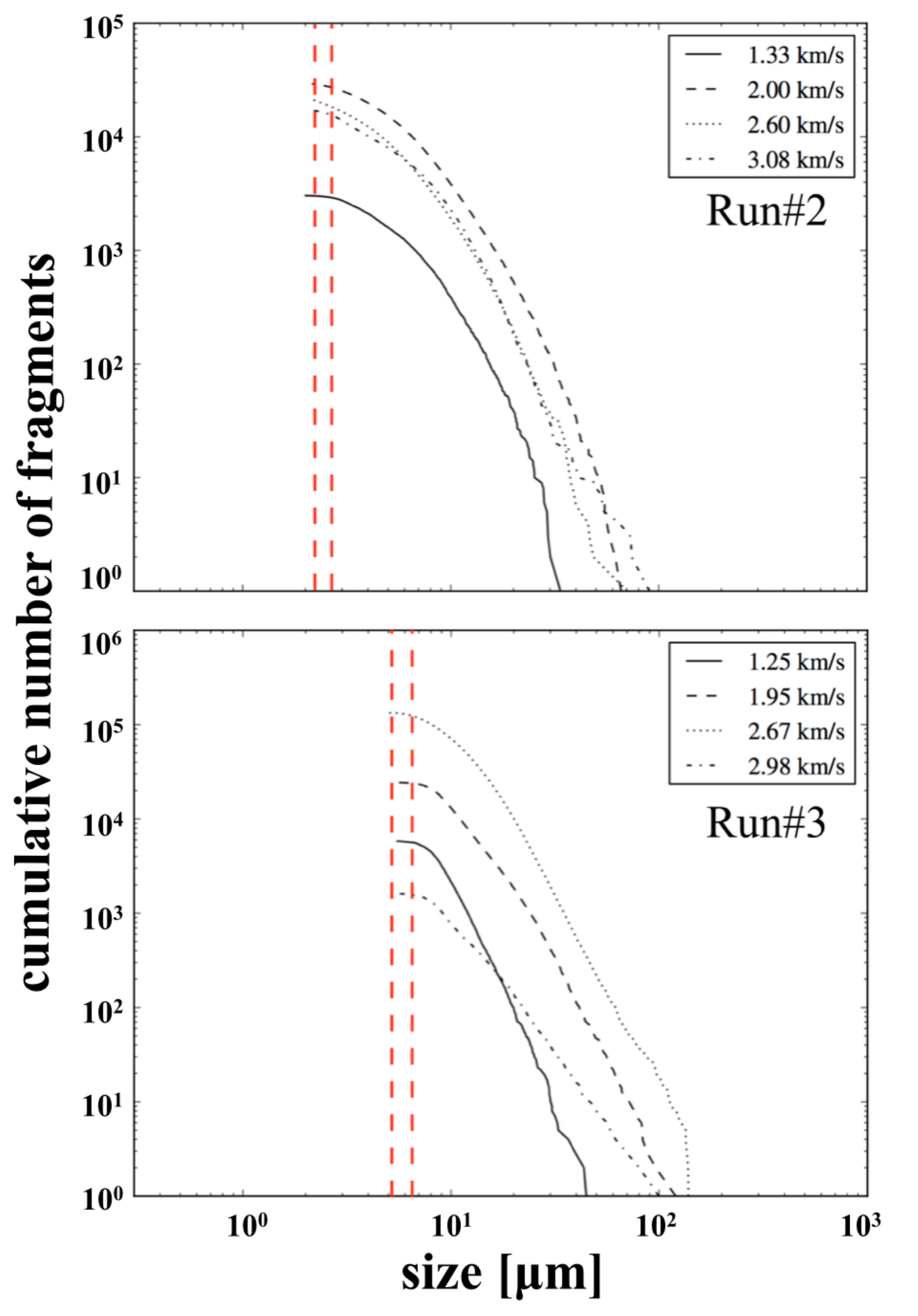}
\caption{SFD of the ejecta fragments of indicative shots in Run\#2 and Run\#3, showing no significant change of the slope with increasing speed in the same Run. The red dashed lines indicate the threshold range of the detection limit which are approximately at 2.2--2.7 $\mu$m and 6.2--7.4 $\mu$m respectively. Different detection limit for different magnification. It is debatable whether these turnovers are real and not due to the resolution limit of the EDX mapping and/or the SEXTRACTOR software cutoff. A small shift towards larger sizes exists in the ejecta SFD for the shot with the lowest impact speed tested in Run\#2, similar result with Run\#1 (usually for the shot with the lowest speed the projectile was recovered completely intact). Note that the largest recovered fragments are not included in this distributions.}
\label{sfd}
\end{figure}

Figure~\ref{sfd} shows that the range of slopes of the SFDs per experimental Run does change with increasing target porosity -- distributions become steeper -- and was calculated to be between $-2.5$ and $-4.0$ for Run\#2 and $-3.0$ and $-4.8$ for Run\#3, consistently higher compared to Run\#1 where they lie between $-1.04$ and $-1.68$ \citep{me2016}. Surprisingly, there is no trend in the slopes of the SFDs at different impact speeds in the same Run. The same result was obtained for olivine projectiles fired onto the non-porous water-ice targets \citep{me2016} implying that the impact speed (up to 3 km/s used here) does not substantially affect the fragmentation behaviour of the peridot projectiles. This result is in contrast to the `common-sense' assumption that the impactor should produce more numerous, and smaller, ejecta fragments - and thus steeper ejecta SFDs - when it hits the same target at higher speeds. One explanation could be that olivine debris underwent secondary fragmentation on the ejecta collecting systems. However, we expect this secondary fragmentation to be limited, due to the low ejecta speeds, which are only a small fraction of the incident speed. Another explanation is that the peridot projectiles have a fragmentation behaviour different from that of more ductile (i.e. metal) projectiles \citep{Hernandez2006,2013kenkmann,mcdermott2016}. 

Nevertheless, there is a clear trend of increasing steepness of the slopes with increasing target porosity. This means that the fraction of the small fragments produced is greater than the larger ones. As the target's porosity increases, the bulk target itself becomes weaker and thus is easier to penetrate. But, on the other hand, the increased porosity makes the target to `be seen' harder from the projectile's perspective. The increased macroporosity, means larger voids inside the target that dissipate the energy that is delivered by the impact. As both porous targets consisted of grains, the impact mechanism was not the same as on a non-porous target. During an impact onto a solid material, a shock-wave is produced and penetrates the target as well as the projectile. At progressively higher speed impacts, a stronger shock-wave will be produced and (depending on the projectile size) will totally penetrate the projectile backwards and when the shock-wave reaches the rear to move again forward and so forth: this causes the fragmentation of the projectile. Whilst impacts on non-porous materials will only `see' one target, in porous targets, comprised of grains, with a size comparable to the projectile (as in Run\#2), multiple impacts may occur as the impactor penetrates the target material. Each of these impacts will cause the production of a new shock-wave. Therefore, the projectile will suffer multiple shock-events, something that will lead to higher fragmentation. 

\begin{figure}
\includegraphics[width=1.\columnwidth]{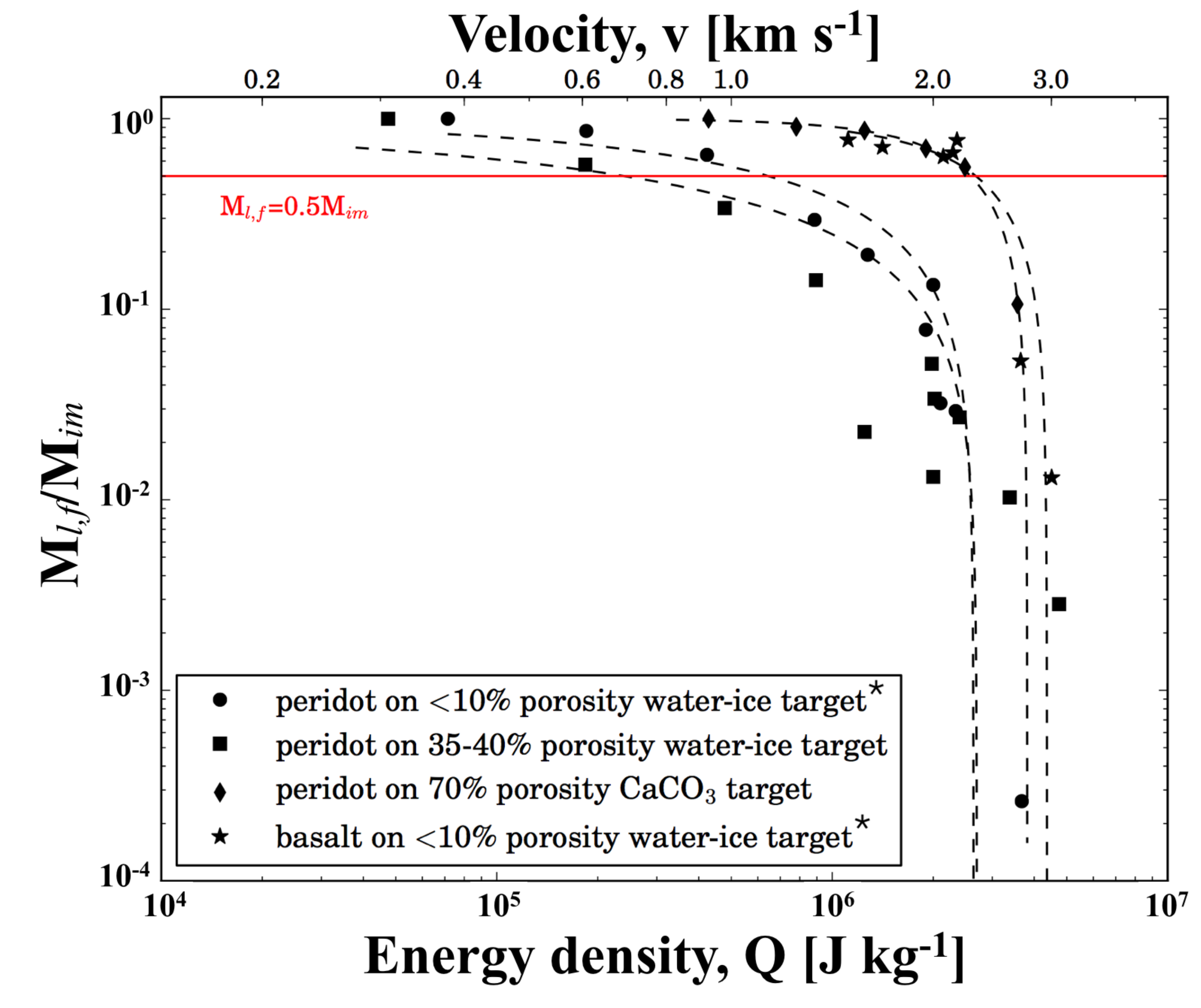}
\vspace{-15pt}
\caption{The largest recovered fragment ($M_{\mathrm{l,f}}$) relative to the initial mass ($M_{\mathrm{im}}$) from all the shot Runs, including data from \protect\cite{me2016}. The dashed lines correspond to the best fitting of a power law. The red solid line indicates the limit for the catastrophic disruption.}
\label{qplot}
\end{figure}

%qplot
Another way to compare the fragmentation of the peridot projectiles during the three Runs was to look for differences in the largest surviving fragments after each shot, and also what speed (or energy) is required for its catastrophic disruption to occur, when M$_{\mathrm{l,f}}$/M$_{\mathrm{im}}$=0.5 (see Table~\ref{qcatastrofic}).
In Figure~\ref{qplot} we present the mass of the largest fragment we retrieved as a fraction of the initial impactor's mass, in relation to the energy density, which, for a given impact speed $\upsilon$ (m/s), is defined as $Q_{\mathrm{im}}=\frac{1}{2} v^{2}$.

The synthetic basalt projectiles, used in Run\#1, have a comparable size to the peridots, but require an order of magnitude higher energy to retain 50\% of their initial mass. So catastrophic disruption for peridot projectiles occurs at 1.14 km/s, whereas for synthetic basalt this happens at 2.33 km/s, indicating that peridots are more fragile than basalt. This is in agreement with the comparison of the compressive strengths of both materials; which are 80 MPa and 100--250 MPa for olivine and basalt respectively \citep{propertiesmeteorites,basaltstrength}. 

There is a small shift of the data towards smaller energies (see Fig.~\ref{qplot}) when the peridots impacted the porous water-ice targets (Run\#2), in comparison with the same projectiles onto the non-porous water-ice targets (Run\#1), and this difference is more obvious at lower speeds. Moreover the tail of the plot for the porous target appears to be less steep: 50\% of the initial impactor mass is preserved at collisional speeds of 1.14 km/s and 0.60 km/s respectively, giving a reduction of the energy density of $\sim$3. Upon increasing the porosity of the target (70\%) when the CaCO$_{3}$ powder was used, we expected to see a further shift of the energies towards lower values, following the same behaviour as stated earlier. However this is not observed. On the contrary, the whole dataset shifts to the right (relative to the data from the non-porous water-ice targets) with the collision speed for catastrophic disruption occurring at 2.27 km/s, where we find the large fragments of the synthetic basalt, giving an increase in the energy density of an order of magnitude (see Table~\ref{qcatastrofic}). 
Therefore target's porosity is not the only parameter for the impactor's fragmentation.
From the shift towards higher energies in Figure~\ref{qplot} for the regolith targets, it is apparent that the target material and the target's material grain size also contribute to the result. In Run\#2, where peridots impacted onto $\sim$40\% porosity water-ice targets, the ice grain sizes were in the size range from a few mm (similar to the impactor's size: 3 mm) down to 10s of microns. While in Run\#3, where peridots hit the regolith CaCO$_{3}$ powder (highest porosity, finest grained), the average grain size dropped significantly to microns, similar to the finest water-ice `grains'.

\begin{figure}
\includegraphics[width=1.\columnwidth]{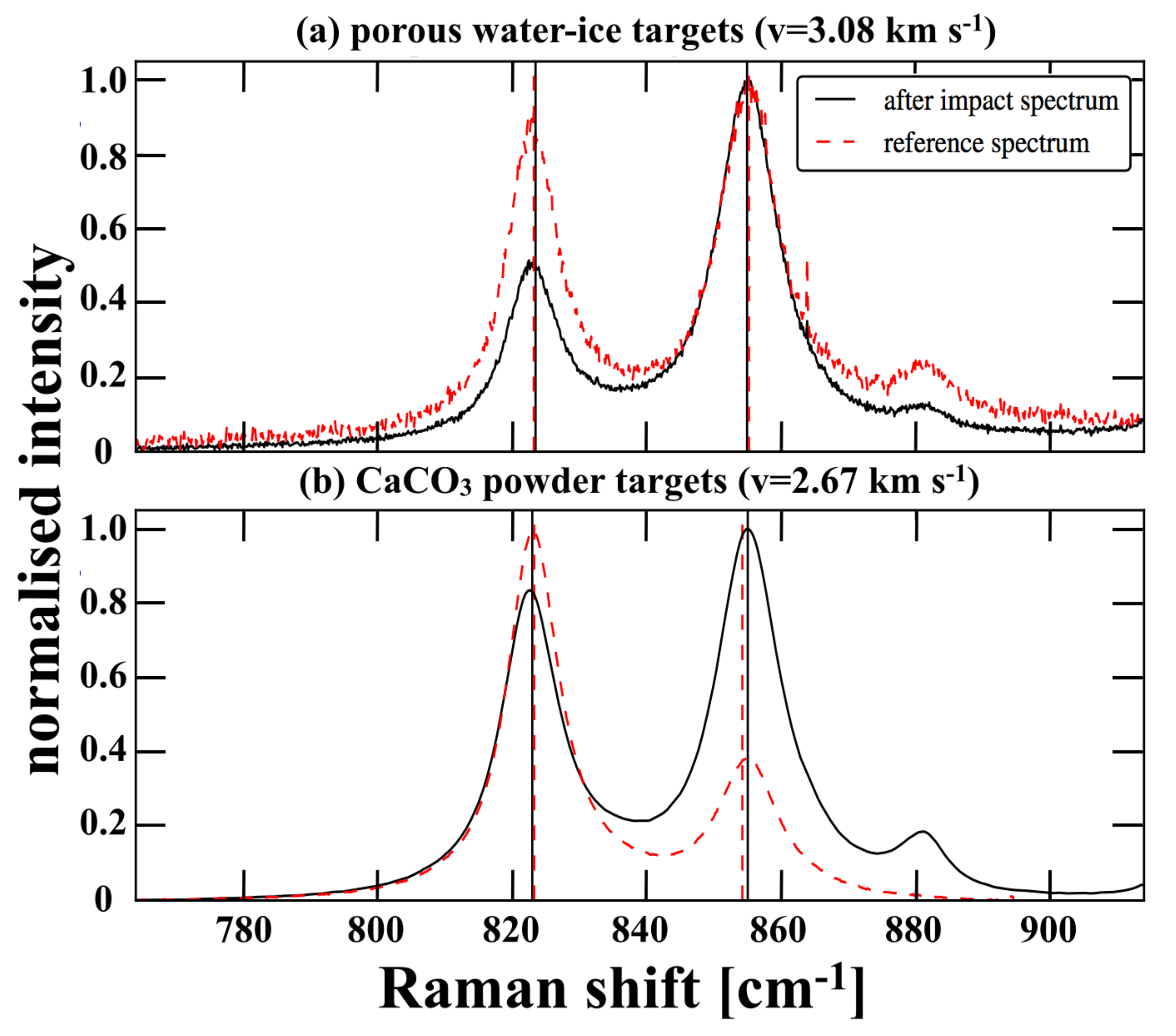}
\vspace{-15pt}
\caption{Raman spectra of large fragments that recovered after shots (solid line) during Run\#2 and Run\#3, in  comparison with the reference (red dashed line). No significant change above the instrument precision was observed in the P$_1$ and P$_2$ olivine lines.}
\label{both}
\includegraphics[width=1.\columnwidth]{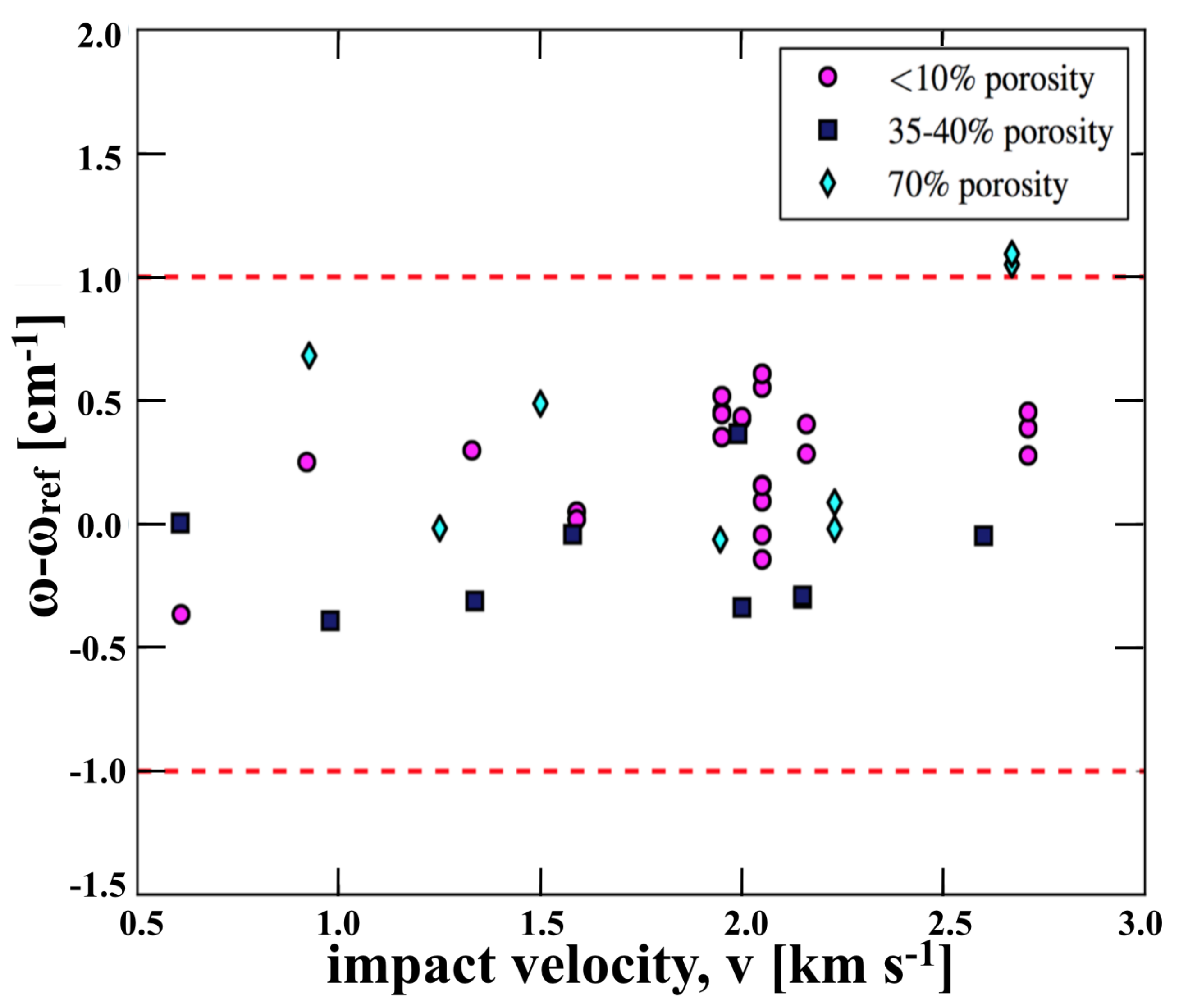}
\vspace{-15pt}
\caption{The change in separation, $\omega$, of the P$_1$ and P$_2$ olivine lines was calculated for all the largest recovered fragments in the range of impact speeds 0.60--3.08 km/s including results from \protect\cite{me2016}. The dashed lines indicate the sensitivity limit of the spectrometer.}
\label{diffIR}
\end{figure}

%raman
Further investigation of all the large fragments, using the Raman spectrometer as detailed in \cite{me2016}, showed no indication of impact melting, as the Raman spectra show no change in the position or mutual distance of the two characteristic olivine peaks P$_{1}$ and P$_{2}$ (Fig.~\ref{both} and \ref{diffIR}). We assume that, up to the tested impact speeds (that gave large identified fragments), the velocity is too low to produce any significant change to the peridot impactor material. The other explanation is that the origin of the material examined under Raman spectrometer was `far' from the impact point (i.e. originated at the middle or back of the projectile) and, thus, was not affected strongly by the impact shock. Inspection of the recovered fragments gave no indication where they were originally located within the projectile.

\begin{table}
\centering
\caption{The energy density at catastrophic disruption limit, for olivine and basalt, after fitting a power law to the experimental data, $M_{\mathrm{l,f}}/M_{\mathrm{im}}$ = 1-AQ$_{\mathrm{im}}^{c}$, where A and c the fitting parameters. **Data from \protect\cite{me2016}.}
\vspace{5pt}
\begin{tabular}{l|ccc}
\hline
\hline
peridot projectile & Q*$_{\mathrm{p}}$ [J/Kg]  & A & c \\
\hline
Run\#1**  & 6.46$\times$10$^{5}$  & 6.89$\times$10$^{-4}$ & 0.49\\
Run\#2  & 2.40$\times$10$^{5}$& 1.4$\times$10$^{-2}$ & 0.28 \\
Run\#3  &2.58$\times$10$^{6}$ & 1.8$\times$10$^{-12}$ & 1.78\\
\hline
basalt projectile & Q*$_{\mathrm{b}}$ [J/Kg]& A & c\\
\hline
Run\#1** & 2.71$\times$10$^{6}$&3.63$\times$10$^{-10}$& 1.42\\
\hline
\hline
\end{tabular}
\label{qcatastrofic}
\end{table}

\subsection{Implantation of material in the target}\label{implantation}

The total mass was estimated as the sum of the mass of the fragments that were directly recovered visually from the target immediately after the shot, and the mass calculated following the same procedure as described above, analysing the filters using the EDX and SEXTRACTOR technique, measuring their x and y dimensions, assuming that the produced fragments are cuboids and have constant density $\rho$=3.18 g cm$^{-3}$. 
In Run\#1 the projectile leaves a few per cent of its initial mass in the targets even at impact speeds $\textgreater$2.0 km/s. In the subsequent Runs, where the porosity is higher, the amount of implanted material increased considerably. It should be mentioned though that only in Run\#3 the embedded mass decreased consistently with increasing impact speed, as was expected. In Run\#1 and Run\#2, where water-ice targets were used, there is no clear trend observed, but the implanted mass fluctuates in the range of tested speeds. This result may be biased up to some extent, as from Run\#3 there is no mass estimation of the very small fragments that remained in the target, as due to the contaminating residue left after dissolving CaCO$_{3}$, it was not possible to perform the EDX mapping. However, the missing is small and the recovered mass of the large fragments found in targets, consisted of a very large fraction of the initial impacting mass. Moreover, as mentioned before, in Run\#2, the fact that the target's grain size was similar to the impactor's, also contributed to the fluctuation of the implanted masses.

From the above, it is clear that the target's porosity plays a definite role also in the degree of implantation of the impactor's material on the target after a collision. The ejecta velocities also decreased as the target's porosity increased. During Run\#1 ejecta flew backwards $\sim$50 cm, but in Run\#3, no ejecta was recovered further more than a few cm ($\sim$5 cm) from the impact point. The low ejection velocities may also have contributed to the non-escape of the projectile's material from the target. For a porous material the ejecta velocities have been measured to be up to two orders of magnitude less than the ejecta speeds measured for rocks. This means that for the impact speeds tested in a laboratory the ejecta cannot fly with a speed beyond a few m/s \citep{2002holsappleast3}. Another contributing factor could be that the largest fragments of the ejecta travel with lower velocities compared to the small ones \citep{1999Icar..142....5B}. Also, this last factor was partly responsible for the majority of the largest fragments of Run\#2 and Run\#3 being recovered directly from the target. It has already been shown in previous experiments \citep{schultz2007} that the projectile penetrate the porous targets in greater depths and this leads to greater retention of its mass.
The effect of the Earth's gravity is an important extra factor in these experiments: as the gun is fired horizontally, loosened material that might otherwise remain in a crater in the case of impact on a minor body, will be lost in our experiments. This is because the Earth's gravity acts to reduce the amount of material remained in the crater when the impact is occurred horizontally. This indicates that our results for the implantation of the impactors material correspond to a lower value.

\begin{figure}
\includegraphics[width=1.\columnwidth]{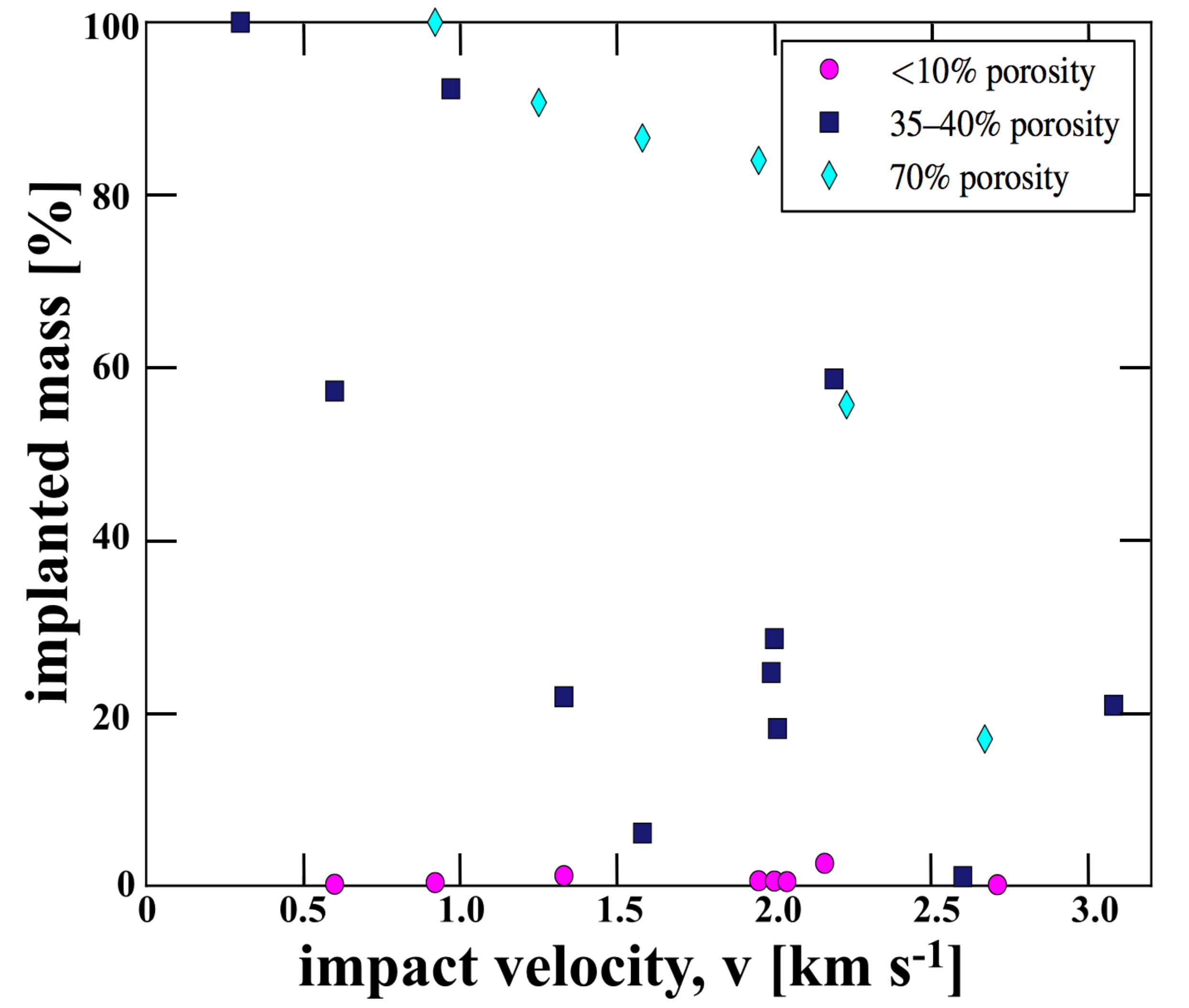}
\vspace{-15pt}
\caption{The implanted projectile's mass on the target significantly increased with increasing target's porosity. However, when the target's grains had comparable size with the projectile (Run\#2), there is no clear trend with increasing speed.}
\label{masses}
\end{figure}
 
\section{Conclusions}\label{conclusions}
We confirm that, in the vertical impacts with 90$^{\circ}$ impact angle, porosity plays a significant role in the fragmentation of the impactor but, more importantly, on the amount of the implanted mass on the target. This result has implications for studies on large-scale collisions between asteroids in Main Belt. Although it was initially believed that the impactor after a high-speed collision is pulverised (and/or vaporised) and not able to embed material into the target body, it is shown herein that such studies should be revised, thus altering the big picture of collisions in the Main Belt, providing formation scenarios for the observed spectral variability of some asteroids or even the formation of multi-lithology objects. Future spacecraft observations of asteroid surfaces and sample-return missions, such as Hayabusa II and OSIRIS-REx, will provide invaluable information also for the collisional history of such bodies. Can we find exogenous material on C--type asteroids? It is hoped that the work presented here will help to interpret the data from such space missions.

\section*{Acknowledgements}
CA would like to thank the UoK for her 50$^{th}$ Anniversary PhD scholarship. MD acknowledges support from the French Agence National de la Recherche (ANR) SHOCKS and the French Programme National de Plan\'etologie (PNP). MCP and MJC thank the STFC, UK for funding this work. 

\bibliographystyle{mn2e}
\bibliography{references}

\label{lastpage}

\end{document}